\documentclass[twocolumn, secnumarabic, amssymb, amsmath,nobibnotes, aps, prl, showpacs]{revtex4}
\begin{document}
\title{Fundamental Inconsistencies of \lq Superstatistics\rq}
\author{B. H. Lavenda}
\email{bernard.lavenda@unicam.it}
\affiliation{Universit\'a degli Studi, Camerino 62032 (MC) Italy}
\date{\today}
\newcommand{\sumi}{\sum_{i=1}^{n}\,}
\newcommand{\sumj}{\sum_{j=1}^{n}\,}
\newcommand{\sumk}{\sum_{i=1}^{k}\,}
\newcommand{\half}{\mbox{\small{$\frac{1}{2}$}}}
\begin{abstract}
Fundamental inconsistencies of superstatistics are highlighted. There is no such thing as a superposition of Boltzmann factors; what is actually derived is a generating function and not a normalizable probability density. The beta density of the second kind for the energy is shown to be subordinated to  $\chi^2$-probability distribution, and this provides a power-tail distribution for large values of the energy, comparable to a Pareto distribution. The authors' claim that the Laplace method will only work on the inverse Laplace transform in the opposite limit of small energies is shown to be incorrect. The exact expression is obtained by contour integration at a multipole located at the origin, corresponding to the high temperature limit. 
\end{abstract}
\maketitle
Notwithstanding the criticisms logged against \lq superstatistics\rq\cite{LD}, the manuscipt \lq Asymptotics of Superstatistics\rq\ has recently been posted on this site \cite{TB}. The basic idea is to obtain \lq fat\rq\ tail power distributions by a \lq mixing\rq\ of Boltzmann factors. The rationale behind why such a mixing of Boltzmann factors will produce fat tail power laws is left to the reader's intuition.\par
A \lq mixing\rq\ distribution, $f(\beta)$, is used to obtain a new \lq distribution\rq, $B(E)$, through a Laplace transform
\begin{equation}
B(E)=\int_0^{\infty}f(\beta)e^{-\beta E}\,d\beta=
\left<e^{-\beta E}\right>, \label{eq:Laplace}
\end{equation}
where $E$ is the energy and $\beta$ is its conjugate intensive parameter. It is claimed that both $f(\beta)$ and $B(E)$ are \emph{unimodal\/}. This is unnecessary and misleading: all that is required is that they be \emph{monotone\/} \cite{Feller}. In the manuscript under discussion, the Laplace integral, (\ref{eq:B}) was evaluated in the high energy, $E\rightarrow\infty$, $\beta\rightarrow0$ limit, using Laplace's  method. Yet, when the inverse transform is considered,
\begin{equation}
f(\beta)=\frac{1}{2\pi i}\int_{b-i\infty}^{b+i\infty}
B(E)e^{\beta E}\,dE, \label{eq:in-Laplace}
\end{equation}
along the vertical line $\Re\beta=b$, in the region of existence of $B(E)$, we are surprised to learn that the same method can only be applied in the opposite limit as $E\rightarrow0$, $\beta\rightarrow\infty$. This appears all the more disconcerting since no information is lost in a Laplace transformation: A distribution is uniquely determined by its Laplace transform and vice-versa. Both should yield the same expressions in the same asymptotic limit. \par
Historically, any relation giving the asymptotic behavior of $B(E)$ in terms of $f(\beta)$ is known as a Tauberian theorem, whereas the inverse relation is referred to as Abelian \cite{Feller}. The behavior of $B(E)$ as $E\rightarrow\infty$  determines completely the behavior of $f(\beta)$ near the origin and vice-versa. Without loss of generality, we can consider power laws for both $f(\beta)$ and $B(E)$, one of which must be monotonically increasing, while the other monotonically decreasing. That the Laplace transform (\ref{eq:Laplace}) exist, it is necessary that the exponent in \begin{equation}
f(\beta)=\frac{\beta^{m-1}}{\Gamma(m)}, \label{eq:f}
\end{equation} be confined to the semi-open interval $[0,\infty)$. This then gives 
\begin{equation}
B(E)=E^{-m}, \label{eq:B}
\end{equation}
which is the tail distribution that Touchette and Beck derive by Laplace's method in the limit as $E\rightarrow\infty$. They are quick to point out that (\ref{eq:B}) has the same asymptotic form as Tsallis' $q$-exponential distributions \cite{Tsallis}, where $m=(q-1)^{-1}$.\par
However, the Laplace transform (\ref{eq:Laplace}) defines a generating function, $B(E)$, and not a normalizable probability density. The conjugate distribution is the $\chi^2$ distribution \cite{BHL04}
\begin{equation}
p(\beta|E_0)=E_0\frac{(\beta E_0)^{m-1}}{\Gamma(m)}e^{-\beta E_0}, \label{eq:chi}
\end{equation}
characterized by the \lq parameter\rq, $E_0$, and not the probability distribution \cite{TB}
\[p(E)=B(E)/Z,\]
where
\[Z=\int_0^{\infty}B(E)\,dE,\]
since the integral diverges. Hence, a simple Laplace transform on the \lq density of states\rq, $f(\beta)$, will not lead to a power-tail distribution.\par
The $\chi^2$-distribution (\ref{eq:chi}) for the inverse temperature can, however, be derived from the $\chi^2$-distribution for the energy by Bayes' theorem of inverse probability \cite{BHL91}. Moreover, it is  the distribution of the  \emph{directing\/} process which transforms the $\chi^2$-distribution 
\begin{equation}
p(E|\beta_0)=\beta_0\frac{(\beta_0 E)^{m-1}}{\Gamma(m)}e^{-\beta_0E}, \label{eq:chi-bis}
\end{equation}
for the energy into a beta distribution of the second kind for the energy. This is accomplished through  \emph{subordination\/} \cite{Feller}, in which  the parameter, $\beta_0$, referred to as the \lq state of nature\rq, and characterizes the heat reservoir, is replaced by the lower limit on the energy distribution \cite{BHL95}. \par
  The two $\chi^2$-distributions, (\ref{eq:chi}) and (\ref{eq:chi-bis}), are related by the L\'evy transformation
\[
\beta_0E=\beta E_0, \]
where the subscripted quantities refer to the heat reservoir. In analogy with Brownian motion, where we can consider a distribution of particles at a given instant in time or, equivalently, consider a single particle and follows its motion in time, the conjugate variables of position and time become those of energy and temperature in thermodynamics \cite{BHL95}. The  \lq pure\rq\ spatial process, where the initial position replaces the time parameter, is the Cauchy distribution, and we say that the Cauchy process is \emph{subordinated\/} to Brownian motion \cite[p. 348]{Feller}. Analogously, in thermodynamics, the beta density of the second kind is subordinate to the $\chi^2$-distribution (\ref{eq:chi-bis}), since \cite[p. 56]{BHL95}
\begin{eqnarray}
p(E|E_0) & = & \int_0^{\infty}p(E|\beta)p(\beta|E_0)\,d\beta\nonumber\\
& = & B^{-1}(m,m)\frac{E_0(E_0E)^{m-1}}{(E_0+E)^{2m}},
\label{eq:beta}
\end{eqnarray}
where $B(\cdot,\cdot)$ is the beta function. For $E_0\ll E$, the beta density (\ref{eq:beta}) transforms into the Pareto density
\[
p(E|E_0)\sim\frac{E_0^m}{E^{m+1}}.\]
The energy $E_0$ is then interpreted as the smallest energy for which the distribution is valid, corresponding to the minimum income in Pareto's distribution of income above a given value.\par
Touchette and Beck claim that Laplace's method will only work on the inverse transform (\ref{eq:in-Laplace}) in the limit as $\beta\rightarrow\infty$ and $E\rightarrow0$, i.e., the low temperature limit.  The distribution they find by using Laplace's method is
\[f(\beta)\sim\frac{e^{\beta E_{\beta}+\ln B(E_\beta)}}{\sqrt{-(\ln B(E_\beta))^{\prime\prime}}}, \]
where the prime means differentiation with respect to the argument which is the implicit solution to
\begin{equation}
\left(\ln B(E)\right)^{\prime}=-\beta.\label{eq:E}
\end{equation}
However, $\beta E+\ln B(E)$ has a \emph{minimum\/} at $E_{\beta}=m/\beta$ and not a maximum, as contended by the authors. Hence, Laplace's method will not work regardless of which asymptotic limit is being considered. It is also surprising that the authors find it necessary to work in the opposite asymptotic limit when discussing the inverse transform. No information is lost upon taking the Laplace transform or its inverse so the two should necessarily correspond to the same asymptotic limit.\par
As is well-known, the inverse Laplace transform (\ref{eq:in-Laplace}) can be performed by contour integration. There is a multipole of order $m$ at the origin, and $B(E)\rightarrow0$ as $E\rightarrow\infty$. If we enclose the pole at the origin by a contour $C_m$, we have from Cauchy's theorem
\begin{eqnarray*}
f(\beta) & = & \int_{C_{m}}E^{-m}e^{\beta E}\,dE\nonumber\\
&= & \frac{1}{\Gamma(m)}\left[\frac{d^{m-1}}{dE^{m-1}}e^{\beta E}\right]_{\beta=0}=\frac{\beta^{m-1}}{\Gamma(m)}, 
\end{eqnarray*}
which is precisely the density (\ref{eq:f}) that we started out with.\par
Having come this far, we may like to inquire into the physical meaning of $f(\beta)$ and $B(E)$. According to the process of subordination, the primary process is the $\chi^2$-distribution (\ref{eq:chi-bis}) for the energy. The \emph{directing\/} process is governed by the inverse temperature distribution (\ref{eq:chi}). This is analogous to randomizing time in Brownian motion where the Brownian motion process for the increments of a Brownian particle is transformed into a L\'evy distribution for the random increments in time.\par Equilibrium statistical mechanics  teaches us that the fundamental quantity is the improper probability distribution, or \lq structure\rq\ function  \cite{Khinchin},
\begin{equation}
\Omega(E)=\frac{E^{m-1}}{\Gamma(m)}, \label{eq:Omega}
\end{equation}
representing the density of states in phase space. The conjugate probability distribution is the exponential law
\begin{equation}
p(E|\beta)=\frac{e^{-\beta E}}{Z(\beta)}\Omega(E), \label{eq:conjugate}
\end{equation}
where the generating function is
\begin{equation}
Z(\beta)=\int_0^{\infty}\Omega(E) e^{-\beta E}=\beta^{-m}. \label{eq:Z}
\end{equation}
The functions $f(\beta)$, (\ref{eq:f}) and $B(E)$, (\ref{eq:B}) refer to the inverse of the generating function, (\ref{eq:Z}), and the inverse of the structure function, (\ref{eq:Omega}), respectively. Hence, they are devoid of any physical content.\par
Khinchin \cite{Khinchin}[p. 87] contends that the conjugate distribution (\ref{eq:conjugate}) can be well approximated by the normal distribution  on the strength of the central limit theorem. Setting
\begin{equation}
\frac{e^{-\beta E}}{Z(\beta)}\Omega(E)=\frac{\exp\left\{-\left(E+(\ln Z)^{\prime}\right)^2\right\}}{\sqrt{2\pi(\ln Z)^{\prime\prime}}},
\label{eq:normal}
\end{equation}
and evaluating it at $(\ln Z)^{\prime}=-\beta$ gives the important result
\begin{equation}\Omega(\bar{E})=\frac{Z(\beta)e^{\beta\bar{E}}}
{\sqrt{2\pi(\ln Z)^{\prime\prime}}},\label{eq:Khinchin}
\end{equation}
where the mean energy, $\bar{E}=m/\beta$ is given by the high temperature, equipartition limit. Now, introducing this explicitly into (\ref{eq:Khinchin}) gives \cite[p. 171]{BHL91}
\[
\Omega(\bar{E})=\frac{\bar{E}^{m-1}m}{\sqrt{2\pi }m^{m+\half}e^{-m}}=\frac{\bar{E}^{m-1}m}{m!}=\frac{\bar{E}^{m-1}}{\Gamma(m)}, \]
which is precisely Stirling's expression for $m!$.\par
The reason for this is that the normal law (\ref{eq:Khinchin}) applies in the limit where the number of degrees of freedom of the system has been allowed to increase without limit. Rather, if we maintain a finite number of degrees of freedom, and identify the conjugate distribution with the $\chi^2$ distribution (\ref{eq:chi-bis}) for the energy, then we obtain the expression for the density of states as
\begin{equation}
\Omega(E)=Z(\beta)e^{\beta E}\times\frac{\beta(\beta E)^{m-1}}{\Gamma(m)}e^{-\beta E}=\frac{E^{m-1}}{\Gamma(m)},\label{eq:lavenda}
\end{equation}
without any approximation. Obviously, if we were to evaluate (\ref{eq:lavenda}) at $E=\bar{E}=m/\beta$, we would again get Stirling's approximation, $m!=\sqrt{2\pi}m^{m+\half}e^{-m}$.\par
It is appropriate, at this point, to ask for the distinction between the $\chi^2$- and normal distributions as far as the processes are involved. This is a matter for order statistics. The probability that one of the $n-1$ molecules has energy falling in the interval from $E_1$ to $E_1+dE_1$ is $(n-1)dE_1/E$. There are then $n-2$ molecules remaining, $m-1$ of which have energies $\le dE_1$ while $n-m-1$ have energies $\ge dE_1$. The probability that $m-1$ have energies $\le E_1$ is $(E_1/E)^{m-1}$, while those  having energies $>E_1$ is $\left(1-E_1/E\right)^{n-m-1}$. The product of the three factors is proportional to the beta density of the first kind:
\[
p(E_1)=\frac{1}{B(m,n-m)E}\left(\frac{E_1}{E}\right)^{m-1}\left(1-\frac{E_1}{E}\right)^{n-m-1}. \]
The mean energy of the molecules with energies $\le E_1$ is simply the fraction $(m/n)$ of the total energy.
If $n$ is so large that we may use the equipartition result for the total energy, $E=n/\beta$, and  let $n\rightarrow\infty$, the beta density of the first kind will transform into the $\chi^2$-distribution, (\ref{eq:chi-bis}). The mean value $\bar{E}_1=m/\beta$ is again given by the equipartition result, which is entirely independent of the, asymptotically infinite, total number of particles, $n$.

\end{document}